\documentclass[onecolumn]{aa}
\usepackage{graphicx}
\usepackage{txfonts}
\begin{document}
   \title{Super Star Clusters in the Blue Dwarf Galaxy UM~462
          \thanks{Based on observations obtained at the ESO 3.6m
           telescope of La Silla}
    }


   \author{L. Vanzi}

   \offprints{L. Vanzi}

   \institute{ESO - European Southern Observatory,
              Alonso de Cordova 3107, Santiago - CHILE\\
              \email{lvanzi@eso.org}
             }

   \date{Received 17 April 2003; accepted 19 June 2003}

   \abstract{
   I present optical observations of the Blue Compact Dwarf Galaxy UM~462. 
   The images of this galaxy show several bright compact sources. 
   A careful study of these sources has revealed their nature of young
   Super Star Clusters. The ages determined from the analysis of the stellar
   continuum and $H\alpha$ are between few and few tens Myr. The total
   star formation taking place into the clusters is about 0.05 $\mathrm{
   M_{\odot}/yr}$.
   The clusters seem to be located at the edges of two large round-like
   structures, possibly shells originated in a previous episode of star 
   formation. The sizes of the shells compare well with the ages of the 
   clusters. Evidence for the presence of an evolved underlying stellar 
   population is found.

   }

   \maketitle
%

\section{Introduction}

The galaxy UM~462 (UGC~06850) is a Blue Compact Dwarf (BCD) of relatively 
low metallicity. Low resolution spectroscopy of this object was obtained 
by the objective prism survey of MacAlpine et al. (1983), higher quality 
data were collected later on by Terlevich et al. (1991). From these spectra
Masegosa et al. (1994) derived a metallicity 12+Log(O/H)=7.98. More recent
observations by Izotov \& Thuan (1998) gave 12+Log(O/H)=7.95, that is 
about 1/9 solar. A radio HI map has been obtained by van Zee et al. (1998), 
they derive a neutral hydrogen mass of $2.65~10^8~M_{\odot}$. UM~462 has not
been detected in the millimetric CO (Gondhalekar et al. 1998). 
The most recent optical images available are those of Cairos et 
al. (2001), they show a quite compact object, about $20\times10$ arcsec in
size, with a double peaked morphology and blue colors, 
B-V=0.6-0.3, V-R=0.3, V-I=0.2-0.4. No further details of the morphology of 
the galaxy can be inferred from these images due to the low angular resolution. 
The galaxy is classified
as a peculiar BCD in the Near-IR Galaxy Morphological Atlas of Jarret
(2000). Vanzi et al. (2002) identify, for the first time,
a complex morphology in UM~462. Their near-infrared images reveal the presence
of at least 6 young compact clusters that, according to the colors and
luminosities observed were tentatively classified as Super Star Clusters 
(SSC). This discovery and the lack of high quality optical images motivated 
the work presented here. The paper is organized as follow; in Sect. 2
the new observations are described, Sect. 3 is dedicated to the 
description of color-color diagrams, Sect. 4 to the properties of the
clusters: extinction, age, luminosity, mass, size and morphology. The 
conclusions are summarized in Sect. 5.

\section{Observations}

The observations presented in this paper have been obtained on April 31, 2002
at the ESO 3.6m telescope of La Silla using the optical imager spectrometer
EFOSC-2. Images in the U, V, R
and gunn-i broad band filters were obtained with total integration times of 
1800, 1350, 900 and 1200 sec respectively. Each image was divided in three 
exposures of equal duration. The data reduction has followed the standard
steps, fringes had to be removed from the gunn-i images using sky frames.
The photometric calibration was performed using a photometric field of
Landolt (1992). A transformation from gunn-i to I, available at the telescope,
has been applied. The most prominent feature of the images is the presence of
several compact sources. The photometry of the sources detected has been 
obtained
over apertures of  1 \arcsec~ in radius. The contribution of the galaxy
has been measured on an anulus around each source and subtracted. For each
image the photometry has been carried out independently, then values 
corresponding to the same filter have been combined, the dispersion of the measures
give an estimate of the photometric errors that are always between 0.01 and
0.02 mag. The results of the photometry are presented in Table \ref{phot}.
In the last column, indicated as G, the photometry of the
entire galaxy, obtained with a radius of 20 \arcsec~, is reported, the values 
are in full agreement with those of Cairos et al. (2000). 

\begin{figure}
\centering
\includegraphics[width=10cm]{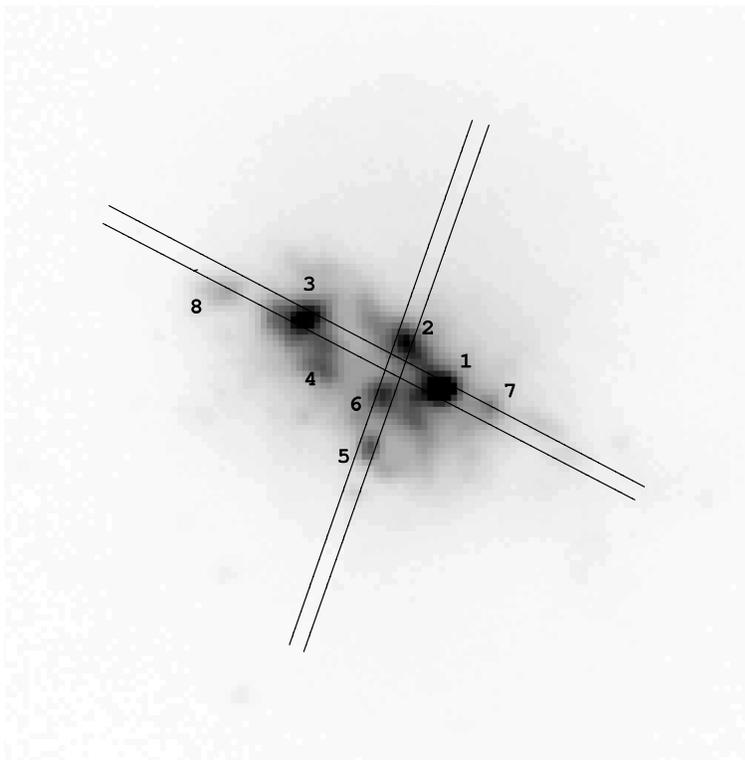}
\caption{Image of UM~462 in R with the clusters indicated by numbers and the
slit positions used for the spectroscopic observations. North is up, east to
the left. The field of view is about $30\times30$ \arcsec.}
\label{r}
\end{figure}

\begin{figure}
\centering
\includegraphics[width=14cm]{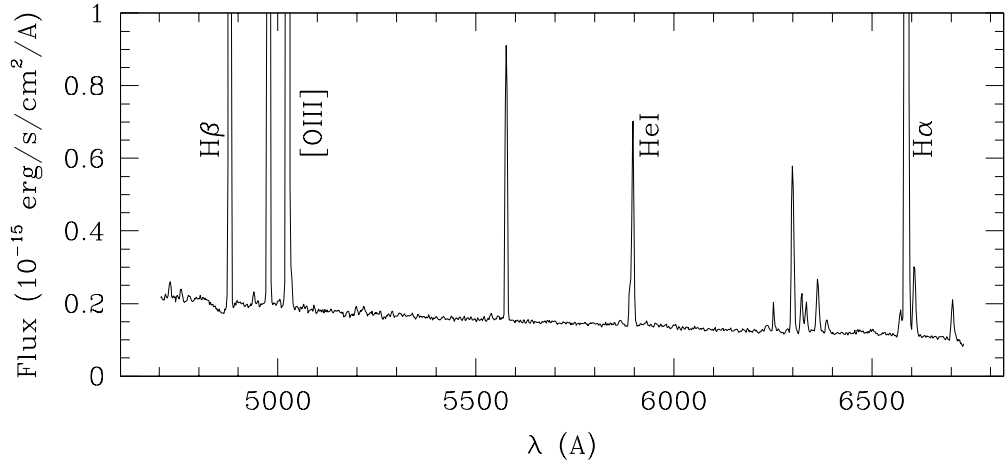}
\caption{Optical spectrum of Cluster 1. The emission lines not labeled
at 5577 and around 6300 \AA~ are from the sky.}
\label{spc}
\end{figure}

\begin{figure}
\centering
\includegraphics[width=10cm]{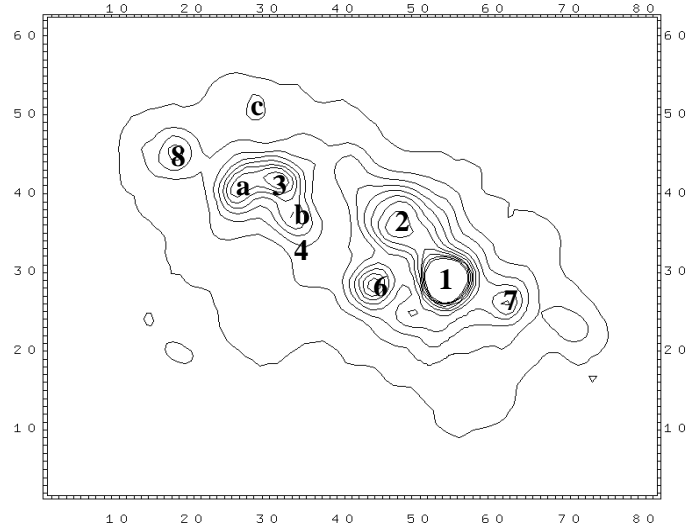}
\caption{Contour image of UM~462 in $H\alpha$. The positions of the clusters are 
indicated by numbers, new sources not visible in the broad band images are 
indicated by letters. North is up, east to the left. 
The field of view is $25\times19$ \arcsec.}
\label{ha}
\end{figure}

Spectroscopy has been obtained with the grism \#9 of EFOSC-2 covering the 
spectral
range 4700-6700~\AA~ with a resolution R=900 and a 1 arcsec wide slit. Two
spectra were observed at position angles which allowed to include 5 of the 
compact sources visible in the images. The orientations of the slits and the 
clusters detected are indicated in Fig. \ref{r} on the R image of the galaxy. 
The reduction of the spectra followed the standard steps, 1D spectra
centered on the compact sources were extracted with an aperture of 2 arcsec. The
flux calibration of the spectra has been obtained using the spectro-photometric
standard star EG-274 (Hamuy et al. 1992, 1994) observed with a 5 arcsec wide 
slit and, in a completely independent way, using the photometry
extracted from the V image. The two methods gave results consistent within
0.1 magnitudes. The spectrum extracted from source 1 is shown in Fig. \ref{spc}.
Several lines typical of HII galaxies are easily identified, the fluxes of the 
lines detected are listed in Table \ref{lines}. 
In addition one image in the $H\alpha$ narrow band filter has been observed
with 2700 sec of integration. The width of the filter is 62 \AA~ allowing the
$H\alpha$, redshifted by about 1000 $\mathrm{Km/s}$, to be covered.
The contribution of the continuum has been 
subtracted using the broad-band R image rescaled by a suitable factor.
The image has been flux calibrated using the $H\alpha$ fluxes measured in the 
spectra. In this way it has been possible to have $H\alpha$ fluxes 
for the sources not covered by the spectroscopic observations, see Table 
\ref{lines}. The accuracy in this case has been evaluated to be of about 10\%. 
Quite interestingly bright sources that are not prominent
in any of the broadband images have been detected in
$H\alpha$, these are indicated with a, b and c in Fig. \ref{ha}.

\begin{table}
\caption{Aperture Photometry (r=1 \arcsec) of the clusters in UM~462 
and corrections for the contribution of the emission lines. The 
clusters are indicated with the nomenclature given in Fig. \ref{r}. G indicates 
the photometry of the whole galaxy obtained with a radius of 20 \arcsec.}  
\begin{tabular}{cccccccccc}
\hline
     &   1   &   2   &   3   &   4   &   5   &   6   &   7   &   8   &   G   \\
\hline
 U   & 17.78 & 18.14 & 17.46 & 19.00 & 18.93 & 18.64 & 18.79 & 18.87 & 14.64 \\
 V   & 17.64 & 18.77 & 18.10 & 19.38 & 19.25 & 18.99 & 19.35 & 19.56 & 14.50 \\
 R   & 17.95 & 18.68 & 18.19 & 19.28 & 19.05 & 19.17 & 19.29 & 19.56 & 14.23 \\
 I   & 18.32 & 18.77 & 18.23 & 19.08 & 18.77 & 19.31 & 19.49 & 19.83 & 14.00 \\
$\Delta$U & 0.13 & 0.07 & 0.05 & 0.02 & 0.01 & 0.06 & 0.07 & 0.05 &  -   \\
$\Delta$V & 0.62 & 0.38 & 0.26 & 0.10 & 0.05 & 0.35 & 0.38 & 0.26 &  -   \\
$\Delta$R & 0.37 & 0.20 & 0.15 & 0.04 & 0.02 & 0.19 & 0.20 & 0.15 &  -   \\
$\Delta$I & 0.00 & 0.00 & 0.00 & 0.00 & 0.00 & 0.00 & 0.00 & 0.00 &  -   \\
\hline
\end{tabular}
\label{phot}
\end{table}

\begin{table}
\caption{Emission lines detected in the clusters. For
the sources indicated by ($\ast$) the $H\alpha$ fluxes have been measured from 
the narrow band image. The fluxes are given in $10^{-15}\ \mathrm{erg/s/cm^2}$.}
\begin{tabular}{cccccccccccc}
\hline
         &  1   &  2   &  3   &4($\ast$) &  5   &  6   &7($\ast$)&8($\ast$) &a($\ast$) &b($\ast$) &c($\ast$)\\
\hline
$H\beta$ & 22.5 &  8.7 &  8.4 &  -  & 0.5  &  6.9 & -  &  -  &  -  &  -  &  - \\
$[OIII]$ & 51.0 & 14.8 & 15.8 &  -  & 0.8  & 12.0 &  - &  -  & -  &  -  &  -  \\
$[OIII]$ &153.0 & 44.1 & 46.9 &  -  & 2.2  & 36.2 &  - &  -  & -  &  -  &  -  \\
HeI      &  3.9 &  2.4 &  2.0 &  -  & 1.3  &  2.5 &  - &  -  & -  &  -  &  -  \\
$H\alpha$& 80.9 & 29.0 & 24.9 & 5.6 & 2.1  & 24.8 & 18.3& 11.9 &20.6& 16.4& 5.4 \\
$[NII]$  &  1.5 &  0.7 &  0.4 &  -  & 0.1  &  0.7 &  - &  -  & -  &  -  &  -  \\
$[SII]$  &  0.8 &   -  &   -  &  -  &  -   &   -  &   - &  -  &-  &  -  &  -  \\
\hline
\end{tabular}
\label{lines}
\end{table}

\section{The color-color diagrams}
\begin{figure}
\centering
\includegraphics[width=10cm]{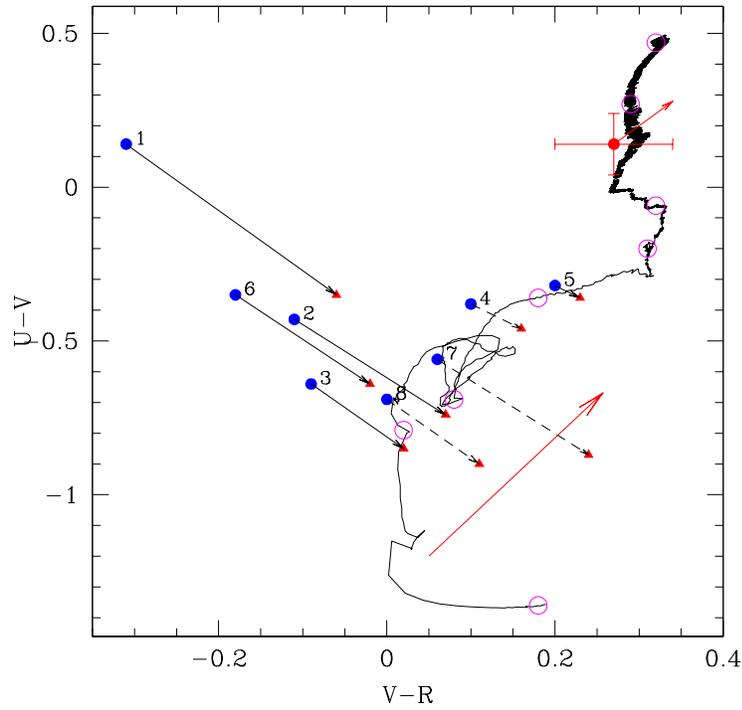}
\caption{U-V/V-R color-color diagram of the clusters in UM~462. Solid circles
are the observed colors, triangles after the correction for the contribution of
emission lines. The dashed arrows indicate when the correction was based on the
$H\alpha$ image rather than on spectroscopy. The dot with error bars 
indicates the colors of the whole galaxy and the arrow the correction for
the clusters\' contribution. The black solid line represents the output of 
SB99 for an instantaneous burst of star formation, points at age 
1, 5, 10, 25, 50, 100,
500 and 1000 $\mathrm{Myr}$ are indicated by open circles. The vector corresponds to
1 mag. of visual extinction.}
\label{uvr}
\end{figure}

\begin{figure}
\centering
\includegraphics[width=10cm]{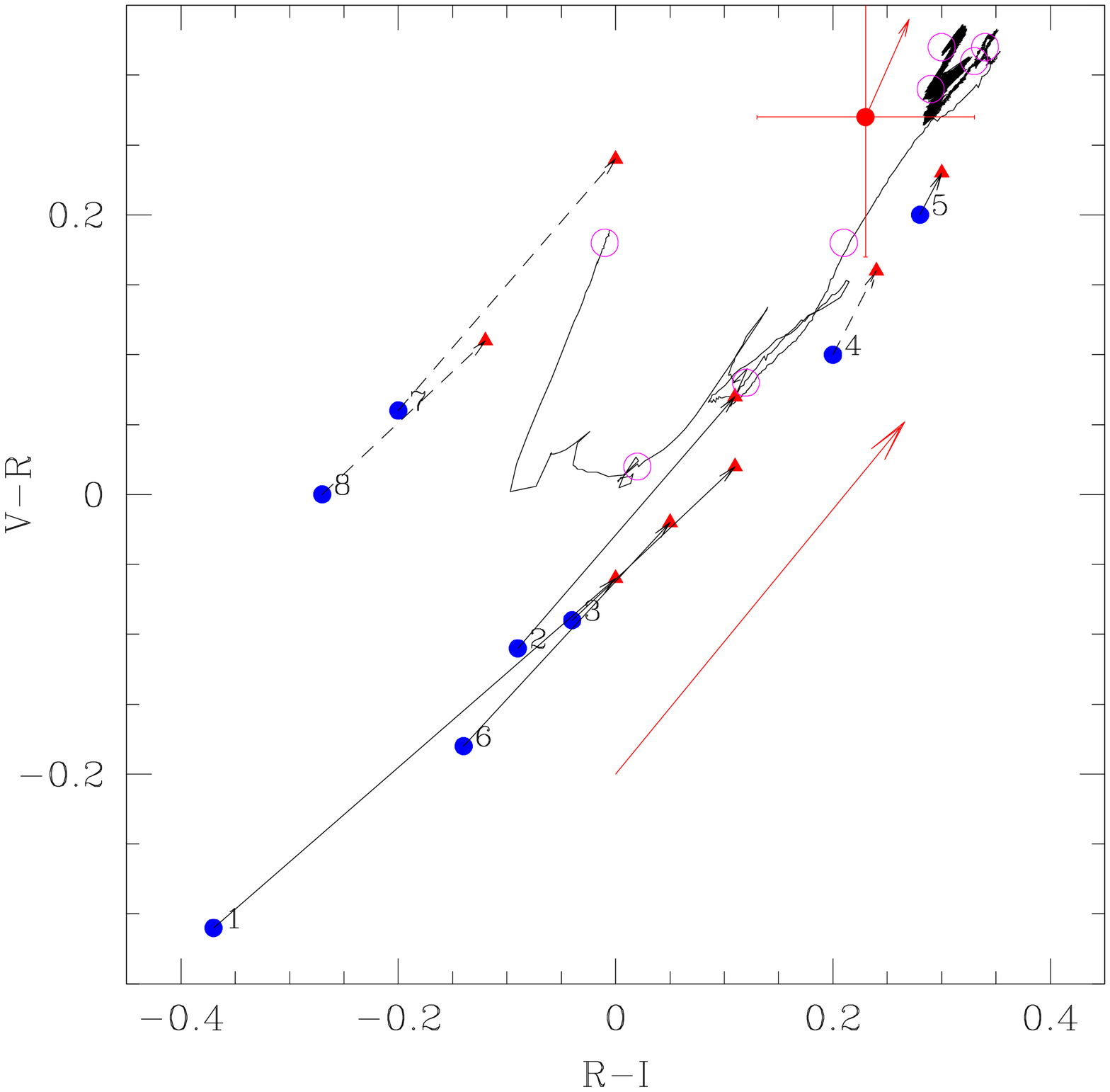}
\caption{Same as the previous Figure for the V-R/R-I diagram.}
\label{vri}
\end{figure}

The sources observed have been studied first comparing their colors
with the evolutionary tracks produced in color-color diagrams by a single 
stellar population model.
To study the stellar component through this method it is critical to evaluate 
any contribution of non stellar origin and to correct for it. The main
non stellar contribution to the broad band magnitudes comes, in our case,
from the gas. The model Starburst99 (SB99 - Leitherer et al. 1999) has been used
as a reference, this model includes the effect of the nebular continuum so 
that it has been sufficient to correct the observations for the contribution 
of the emission lines. This can be done directly through the observations.
Unfortunately the spectra observed do not cover the entire spectral
range which includes the photometric filters used, so that a
direct measure of all emission lines affecting the photometry has not 
been possible. Instead I have used the template spectrum of a typical BCD 
galaxy (Saviane 2003) rescaled and diluted to match the characteristics of the 
spectra available for UM~462 in the region of overlap. 
In particular it has been imposed that the EW of $H\alpha$ and the V magnitude
were the same as those observed. 
These rescaled spectra have been used to measure the contribution of 
the emission lines to the photometry. Of course this approach must be 
considered as not very accurate since the flux emitted in most of the lines 
of interest depends on the temperature, density and abundance; parameters that 
have not been matched perfectly, however the method gives a very good 
idea of the effect and it is certainly the best we can do with the data 
available. 
For those sources not observed in spectroscopy the $H\alpha$ flux obtained
from the narrow band image has been used, the corrections in these cases
must be considered less accurate.
The corrections to the photometry calculated in this way are listed in Table 
\ref{phot}. The effect is most important in V while it is always negligible 
in I due to the absence of important emission lines. 
In Fig. \ref{uvr} and \ref{vri} color-color diagrams for the clusters in 
UM~462 are shown. Both the observed (solid circles) and the
emission-line-corrected points (solid triangles) are plotted. The errors 
on the colors are 0.03 at most.
The tracks generated by SB99 for an instantaneous burst with solar
neighborhood IMF (1-100 $M_{\odot}$, 2.35) and $1/5~Z_{\odot}$ abundance
are plotted as 
solid lines. Despite the uncertainty the emission line correction works very 
well to move the observed points close to the tracks defined by the model. 
On those tracks the points at ages 1, 5, 10, 25, 50, 100, 500 and 1000 
$\mathrm{Myr}$
are marked by open circles with the oldest ages at the top-right
of the plots. 

\begin{figure}
\centering
\includegraphics[width=10cm]{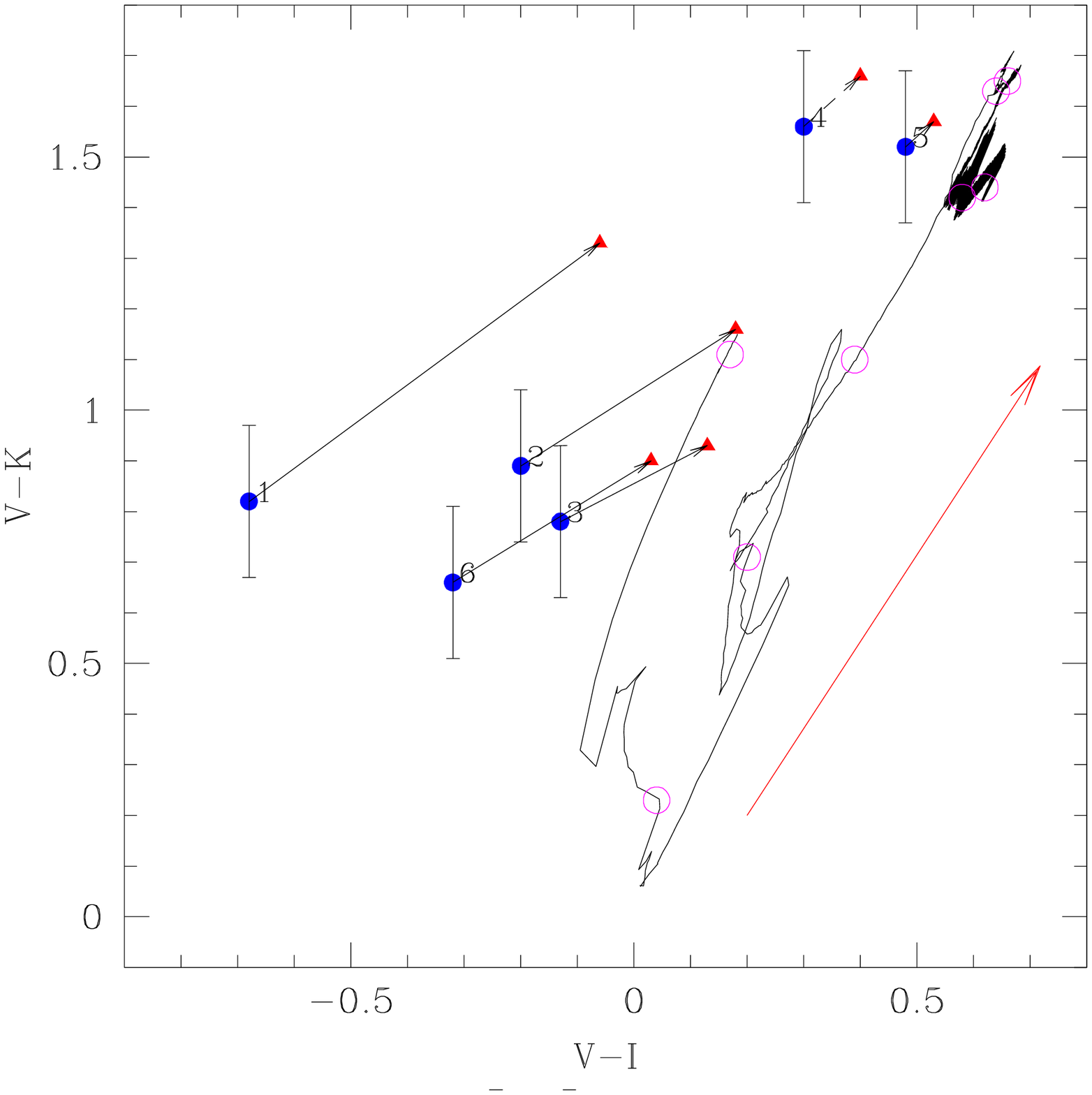}
\caption{Same as the previous Figure for the V-K/V-I diagram.}
\label{vik}
\end{figure}

The near-infrared data from Vanzi et al. (2002) have been used to build
an optical-near-infrared color-color diagram of the clusters. This is shown
in Fig. \ref{vik}. The errors
are larger in this case and the correction for the emission line
contribution more uncertain than in the optical case. The near-infrared
spectrum of Vanzi et al. (2002) in fact includes clusters \#1
only. To correct the K photometry of the other clusters I have used the 
infrared spectrum of cluster \#1 as template and rescaled it to produce for
each cluster the $H\alpha$/$Br\gamma$ ratio predicted by case B, then used
the rescaled spectrum to derive the photometric correction.

The colors of the galaxy as a whole are in all diagrams consistent with the
presence of an evolved population of stars, the observed colors are plotted
as solid circles with error bars, the arrows points to the colors corrected
for the contribution of the clusters, they are consistent with a population
older than 0.5 $\mathrm{Gyr}$. 
Unfortunately the K image is not deep enough to extract a reliable magnitude
for the entire galaxy.

\section{Properties of the Clusters}
\subsection{Extinction}
The extinction can be estimated by the ratio $H\alpha/H\beta$ for the clusters 
with spectroscopic observations. Guseva et al. (2000) give $A_\mathrm{V}=0.27$.
From the fluxes of Table \ref{lines} values in the range 0.11-0.77 can be
derived for cluster \#1, \#2, \#3 and \#6, in particular the highest value of
extinction is obtained for cluster \#1 and \#6. The values of the extinction
are listed in Table \ref{SSC}. For cluster \#1 the $Br\gamma$
flux of Vanzi et al. (2002) can be used in combination with $H\alpha$,
in this case $A_\mathrm{V}=0.55$.  So that the optical extinction can be considered 
relatively low toward all the sources observed. The contribution of the 
galactic extinction in the direction of UM~462, as reported by NED, is low 
$A_\mathrm{V}=0.064$. 

\subsection{Ages}
The stellar populations of all clusters are quite young according to their
colors, typically younger than 10 $\mathrm{Myr}$, as can be derived from the 
comparison with SB99. Clusters \#4 and \#5 are more evolved
than the others. The diagram U-V/V-R is quite straightforward. Clusters
\#1, \#2, \#3, and \#6 show ages of about 5 $\mathrm{Myr}$. 
The gas correction is most 
likely under-estimated for cluster \#1, while it could be over-estimated for
clusters \#7 and \#8 which slightly deviate from the general trend. These
latter clusters could also be very young objects 
affected by some extinction. It is difficult to disentangle the effect 
of extinction and age specially given the uncertainty on the corrections 
applied. We have seen in the previous Section that there are good indications 
for the extinction to be low in UM~462, unfortunately no data are available to 
measure the extinction toward cluster \#7 and \#8.
Clusters \#4 and \#5 are close to the 25 $\mathrm{Myr}$ point of the model.

In the V-R/R-I diagram most of the points seem to show an offset of about 0.1 
mag with respect to the model, the offset however is not systematic and in 
particular clusters \#7 and \#8 present a trend opposite to the others so that 
it is difficult to give a unique interpretation. The problem could arise both 
from the observations and from the model. In any case clusters \#1, \#2, \#3
and \#6 are all close to 5-10 $\mathrm{Myr}$, clusters \#7 and \#8 again deviates from 
the behavior
of the other clusters and are in the very young part of the diagram, 
while clusters \#4 and \#5 are close to 25 $\mathrm{Myr}$ with \#4 a bit 
younger than \#5 as it is in the previous diagram. The optical color-color
diagrams then give fully consistent results.

In the V-I/V-K diagram the data points show a significant offset from the 
model with a color excess of about 0.2-0.5 in V-K. This in principle could be 
due to a systematic error in the V-K color, however the photometry has
been checked with the optical and K photometry available in the literature
both for the galaxy and for the field stars finding good agreement within
the errors. Since the contribution of the underlying galaxy has been 
subtracted and most of the clusters have quite young ages, it seems also 
unlikely for the V-K color to be reddened by the contribution of an evolved 
population. It is interesting to notice that a discrepancy similar to the one 
reported here is also found in the V-I/V-K diagram discussed by Vanzi et al. 
(2000) for the metal deficient galaxy SBS~0335-052 so that it could be due to
a more fundamental effect.
It is well known that current stellar evolutionary tracks fail in reproducing
the correct Blue to Red supergiants ratio (B/R) at metallicities lower than 
solar (Langer \& Maeder 1995, Maeder \& Meynet 2001). In particular the number 
of red supergiants is highly underestimated at low metallicity. 
According to the calibration of Eggenberg et al. (2002) at solar metallicity 
B/R=3 while for $1/5Z_{\odot}$ B/R=0.27. This is mainly due to the longer 
duration of the red phase during the He burning. While solar single stellar 
population models are dominated by red supergiants stars from 6.5 to about
25 $\mathrm{Myr}$ with a maximum at 10 $\mathrm{Myr}$, subsolar model don't. Theory and observations
agree pointing to an even longer red supergiants phase at subsolar abundance.
This has been quantified by Origlia et al. (1999) unfortunately these authors
did not model the V-I and V-K colors, however the evolution of J-K which they
calculate is quite
significant. Standard models predict a maximum in J-K around 10 $\mathrm{Myr}$ which 
rapidly declines at 13-15 $\mathrm{Myr}$, the maximum is much more prominent in solar
than in subsolar models. Adjusting the red supergiants temperature and lifetime
during the He burning makes J-K redder and almost constant beyond 25 $\mathrm{Myr}$.
Since all clusters observed are in the right age interval the effect described 
could well explain the discrepancy of the data respect to the model. 
Despite the systematic deviation from the model again clusters \#1, 
\#2, \#3 and \#6 are close to each other in the young part of the diagram 
while clusters \#4 and \#5 show more evolved colors.
It is interesting to notice that, beside K, the effect must be visible in
the I filter so that it can possibly be responsible for the
red eccess observed for some of the clusters in the V-R/R-I diagram.

The age of the clusters can also be estimated in an independent way by 
the 
comparison of the equivalent width (EW) of $H\alpha$ with the predictions of 
SB99. The ages
derived with this method are listed in Table \ref{SSC}. Clusters 
\#1, \#2, \#3 and \#6 have ages between 4.7 and 5.8 $\mathrm{Myr}$ which are fully
consistent with the values derived from the stellar continuum. Cluster
\#5 instead looks significantly younger in $H\alpha$ than in the continuum
and right at the maximum of the red supergiants phase, a fact that could 
explain the discrepancy for what has been said before.
For cluster \#1, the only one with a NIR spectrum, 
the age can also be derived from $Br\gamma$, the 
value obtained is 4.9 $\mathrm{Myr}$, fully consistent with the previous estimate.  
From the previous analysis no obvious correlation can be found between the
age of the clusters and their extinction.

The sources detected in $H\alpha$ which are faint in the continuum
must have a very high EW, they can easily be clusters younger than about 
4.0 $\mathrm{Myr}$. For these ages in fact SB99 predicts Log EW $(H\alpha) > 10^3 \AA$.
The $H\alpha$ luminosities of these sources are in the range 
$1.4 - 5.4 ~10^{38}\ \mathrm{erg/s}$ that is also compatible with them being
SuperNova Remnants younger than about 2500 days (Turatto et al. 1993).
This latter hypothesis seems less likely since there are no reports of SN
found in the galaxy during the last 15 years.

It must be noted that the tracks used for deriving the
ages of the clusters both from the colors and from $H\alpha$ have a metallicity
slightly higher than UM~462. We can use the tracks at 1/20 $Z_{\odot}$
to qualitatively estimate the effect. The differences introduced even at
this very low metallicity are small and certainly negligible for the present
discussion. 

\subsection{Luminosity and Mass}
Billett et al. (2002) define as Super Star Clusters (SSC) those clusters 
with $M_V < -10.5$ at a fiducial age of 10 $\mathrm{Myr}$. Such an age seems a
good approximation for most of the clusters observed in UM~462.
Using a distance module 30.92 (Vanzi et al. 2002) and the V magnitude 
observed, all clusters qualify as SSCs in agreement with the finding of 
Vanzi et al. (2002). 
If the clusters were younger than 10 $\mathrm{Myr}$, as it is possibly the case for
cluster \#1, \#2, \#3 and \#6, the error introduced would be small, less than
about 0.5 mag, and in both directions as evidenced by the evolution 
of $M_V$ with time calculated by SB99. If instead the clusters were older 
than 10 $\mathrm{Myr}$, as cluster \#4 and \#5 are, then the brightness derived 
would have to be
considered as a lower limits since $M_V$ increases steeply with time.
The number of compact sources observed in UM~462 is very high compared
to the size of the galaxy. If we use the specific frequency defined by
\"{O}stlin (2000) $S_{11}=N_{11} \times 10^{0.4(M_{V,host}+15)}$ where $N_{11}$
is the number of objects with $M_V \le -11$ and $M_{V,host}$ is the total
V absolute magnitude of the host galaxy we obtain $S_{11}=2.46$ which is
the highest value when compared to the list of \"{O}stlin.

The total star formation taking place in the clusters can be derived
from the $H\alpha$ flux, summed over the sources detected. Using the 
calibration of Kennicutt (1998) the total star formation rate is 0.05 
$\mathrm{M_{\odot}/yr}$. This value is in perfect agreement with the star formation
rate of 0.06 $\mathrm{M_{\odot}/yr}$ derived by Vanzi et al. (2002) on the basis
of the IRAS fluxes confirming the low value of the extinction observed and
also telling us that most of the star formation in UM~462 occurs in the 
clusters observed. The value instead contrasts with the 1.4 
$\mathrm{M_{\odot}/yr}$ that
are obtained using the total $H\alpha$ flux measured by Guseva et al. (2000).
The total $H\alpha$ flux summed over all the point like sources detected is
$2.4~10^{-13}\ \mathrm{erg/s/cm^2}$ which compares well with the total flux of
$3.2~10^{-13}\ \mathrm{erg/s/cm^2}$ measured by Terlevich et al (1991), while the 
value of Guseva et at. (2000) is higher by more than one order of magnitude.
The mass can be tentatively derived comparing the luminosity of each
cluster, corrected for the extinction, with the value predicted by SB99 for 
the corresponding age. In deriving the mass of the clusters the IMF of SB99 
that is truncated at a lower mass of 1 $M_{\odot}$  has been extended
to a more realistic lower cut-off of 0.1 $M_{\odot}$. The masses calculated 
in this way are listed in Table \ref{SSC}. For cluster \#5 that has a quite
uncertain measure of the extinction, due to the low fluxes in $H\alpha$ and
$H\beta$ the same extinction as for clusters \#1 and \#6 has been assumed.

\subsection{Size and Morphology}
The observations have been obtained under and average seeing of 1 \arcsec~ in 
V and 0.8 \arcsec~ in I. All clusters are barely resolved in the images
indicating sizes of the order of few 10 pc. Only cluster \#3 is well resolved
with a deconvolved size of 4.8 \arcsec or more than 300 pc. This
indicates that rather than single star clusters we could be observing large
associations, in fact typical sizes for SSCs are of the order of few parsecs
(O'Connell et al. 1995). This also means that the specific frequency derived
in the previous section must be considered as an upper limit.

It is very interesting to notice that the main SSCs observed are distributed 
around two ring-like structures.
One of them, delimited by clusters \#5 and \#6 on the north-est side, is 
quite regular with a radius of about 1.5 arcsec corresponding to 100 pc. 
The second one, surrounded by clusters \#1, \#2, \#3, \#4 and \#6, is more 
elongated with semiaxis of about 2.5 and 4 arcsec equivalent to 166 and 266 
pc. Structures
like these are not unusual in BCD galaxies, they are typically interpreted
as super-bubbles driven by SNe generated in previous SB episodes 
(Martin 1998). The typical expansion velocities measured for these bubbles 
are of the order of few 10 $\mathrm{Km/s}$ (Martin 1998). In the case of UM~462 an 
expansion velocity of 10 $\mathrm{Km/s}$ 
would give an age of about 10 $\mathrm{Myr}$ for the smallest bubble and 
about 20 $\mathrm{Myr}$
for the largest one. These times are perfectly consistent with the idea that 
the current star formation and morphology of UM~462 are the consequence of 
the compression of the ISM produced by a previous episode of star formation.
In this case in fact we expect the expansion age of the bubbles to be greater
than the age of the clusters as it is indeed observed. 

\begin{table}
\caption{Properties of the Clusters, the age is derived from the EW of
$H\alpha$, $A_V$ from the Balmer decrement.}
\begin{tabular}{cccccc}
\hline
     & $M_V$ & EQW($H\alpha$) & age (Myr) & $A_V$ & M($10^5 M_{\odot}$) \\
\hline
  1  & -13.28 & 730$\pm$20 & 4.7 & 0.77 & 7.2  \\
  2  & -12.15 & 325$\pm$20 & 5.1 & 0.49 & 1.2  \\
  3  & -12.82 & 230$\pm$10 & 5.8 & 0.11 & 2.1  \\
  4  & -11.54 &     -      &  -  &  -   &  -   \\
  5  & -11.67 &  33$\pm$ 5 &10.0 &  -   & 2.9  \\
  6  & -11.93 & 260$\pm$10 & 5.6 & 0.77 & 1.6  \\
  7  & -11.57 &     -      &  -  &  -   &  -   \\
  8  & -11.36 &     -      &  -  &  -   &  -   \\
\hline
\end{tabular}
\label{SSC}
\end{table}


\section{Conclusions}

   \begin{enumerate}
      \item High quality optical data of the BCD galaxy UM~462 have been 
       presented. The broad band images reveal several compact
       sources and two large bubbles. A narrow band image in $H\alpha$
       evidences the presence of further compact sources.
      \item Their luminosities qualify all compact sources observed as 
       young Super Star Clusters or very large HII associations.
       The colors of the stellar continuum and the EW of 
       $H\alpha$ and $Br\gamma$ all give fully consistent ages between few 
       and about 25 $\mathrm{Myr}$. The masses are in the range $1.2-7.2~10^5~M_{\odot}$
      \item The number of clusters observed, once normalized to the
       luminosity of the galaxy, is very high, even compared to the most 
       extreme cases known.
      \item The extinction observed in UM~462 is low. Most of the star 
       formation occurring in the galaxy is concentrated in the clusters observed 
       adding up to a total value of $\mathrm{0.05 M_{\odot}/yr}$.
      \item The galaxy as a whole has evolved colors indicating the 
       presence of an underlying evolved stellar population older than
       0.5 $\mathrm{Gyr}$.
      \item The bubbles observed are interpreted as produced by the 
       compression of the ISM due to a previous SB episode. The current
       star formation would have been triggered by this compression. The
       size of the bubbles and the ages of the SSCs are consistent 
       consistent with this idea .
   \end{enumerate}

\begin{acknowledgements}
I wish to thank Luz-Marina Cairos for making her images of UM~462 available 
to me. I am grateful to Michael Sterzik for support during the observations,
to George Hau for providing useful information during the reduction of
the data and to Valentin Ivanov, Marc Sauvage and Andr\'e Maeder for useful 
discussions during the preparation of this paper. Luigi Guzzo provided the 
data to remove the fringes in the i band images. Finally I thank the 
director of La Silla, Jorge Melnick, for granting the observation time for 
this work and the anonymous referee for useful comments which contributed
to improve the paper.

This research has made use of the NASA/IPAC Extragalactic Database (NED) which
is operated by the Jet Propulsion Laboratory, California Institute of 
Technology, under contract with the National Aeronautics and Space
Administration. 
\end{acknowledgements}

\end{document}